\documentclass[10pt]{article}
\usepackage[OE]{express}
\usepackage{graphicx}
\usepackage{epsfig}
\usepackage{epstopdf}
\usepackage{subfigure}
\usepackage{appendix}
\usepackage{amsmath,xfrac,amssymb,mathtools}
\begin{document}
\title{Magnetometry via spin-mechanical coupling in levitated optomechanics}

\author{Pardeep Kumar,\authormark{*} and M. Bhattacharya}

\address{\authormark{} School of Physics and Astronomy, Rochester Institute of Technology, 84 Lomb Memorial Drive, Rochester, New York, 14623, USA}

\email{\authormark{*}kxpsps@rit.edu}

\begin{abstract}
We analyze magnetometry using an optically levitated nanodiamond. We consider a configuration where a magnetic field gradient couples the mechanical oscillation of the diamond with its spin degree of freedom provided by a Nitrogen vacancy center. First, we investigate measurement of the position spectrum of the mechanical oscillator. We find that conditions of ultrahigh vacuum and feedback cooling allow a magnetic field gradient sensitivity of 1 $\mu$Tm$^{-1}$/$\sqrt{\mbox{Hz}}$. At high pressure and room temperature, this sensitivity degrades and can attain a value of the order of 100 $m$Tm$^{-1}$/$\sqrt{\mbox{Hz}}$. Subsequently, we characterize the magnetic field gradient sensitivity obtainable by maneuvering the spin degrees of freedom using Ramsey interferometry. We find that this technique can offer photon-shot noise and spin-projection noise limited magnetic field gradient sensitivity of 100 $\mu$Tm$^{-1}$/$\sqrt{\mbox{Hz}}$. We conclude that this hybrid levitated nanomechanical magnetometer provides a favorable and versatile platform for sensing  applications.
\end{abstract}

\ocis{(270.0270) Quantum Optics; (120.4880) Optomechanics; (350.4855) Optical tweezers; (020.7010)  Laser trapping.} 
 

\section{Introduction}
Recent developments in cavity opto- and electro-mechanics have actively explored the cooling and manipulation of mechanical motion of nano resonators \cite{kippenberg2008,marquardt2009,teufel2011,chan2011,meystre2013}. The spectacular progress in this research has been motivated by ideas for applications in ultrasensitive metrology \cite{aspelmeyer2014} and for testing macroscopic quantum mechanics \cite{armour2002}. However, experimental investigations in this field suffer from heating and decoherence produced by mechanical clamping of the oscillator. An effective remedy to this problem is to isolate the mechanical oscillator from its environment by means of \textit{levitation} using optical \cite{chang2009,neukirch2015,zyin2013} or magnetic fields \cite{isart2012,cirio2012}, for example. In particular, optically levitated nanoparticles have been investigated for cooling \cite{li2011,rodenburg2016,jain2016}, preparation of quantum superposition states \cite{isart2010,isart2011} and precision sensing applications \cite{moore2014,ranjit2015}.

Interestingly, optical levitation of solid state quantum emitters possessing internal degrees of freedom such as Nitrogen-vacancy (NV)-centers provides a promising platform for optomechanical studies \cite{levi2015,hoang2016,pettit2017}. Such hybrid nano-mechanical systems 
are of significant interest for generating macroscopic quantum superposition states \cite{yin2013,scala2013,wan2016,hou2016} and testing quantum gravity in vacuum \cite{andreas2014}. The key ingredient in these proposals is the coupling between the nanomechanical oscillator and spin degrees of freedom mediated by either a magnetic field gradient (MFG) \cite{arcizet2011,kolkowitz2012} or a lattice strain field \cite{ovartchaiyapong2014,golter2016}. Such couplings have been extensively explored for mechanical cooling \cite{rabl2009,zhang2013}, mechanical squeezing \cite{ge2016} and spin squeezing \cite{bennett2013,xia2016}. Due to their easy controllability, fast manipulation, as well as unprecedented coherence time under ambient conditions, NV centers are attractive quantum systems for applications in quantum information processing and ultrasensitive magnetometry \cite{rondin2014}.

More recently, optomechanical system based on whispering gallery modes have been used to realize magnetometers \cite{forstner2012,forstner2014} with magnetic field sensitivity 131 pT Hz$^{-1/2}$ \cite{yu2016}. These magnetometers are miniaturized and offer large dynamic range, unlike superconducting quantum interference devices (SQUID) \cite{romalis2011} and spin exchange relaxation free (SERF) magnetometers \cite{budker2000} which however have better sensitivity. All these magnetometers demonstrate sensitivity to a magnetic field. However, the investigation of MFG sensitivity is equally important due to its potential applications in the generation of macroscopic superposition states \cite{yin2013,scala2013,wan2016,hou2016}, testing of quantum gravity \cite{andreas2014}, magnetic resonance imaging \cite{andreas2007}, and cell therapy \cite{boris2007}. In recent years, non-levitated cantilever type of systems have been used for the measurement of MFG ($\sim 10^{5}$ Tm$^{-1}$ in \cite{kolkowitz2012,rugar2004}).  However, the analysis for the measurement of MFG \textit{sensitivity} in levitated systems still needs to be explored. In this paper, we propose the exploration of MFG sensitivity based on an NV-center levitated in an optical dipole trap. We consider two ways in which the measurement of a MFG sensitivity can be carried out - (i) by manipulating the mechanical degrees of freedom, and taking a trace over the spin degrees of freedom, and (ii) by manipulating the spin degrees of freedom and tracing over the mechanical degrees of freedom.  As recently predicted, maneuvering of the mechanical degrees of freedom under ultrahigh vacuum is expected to lead to the preparation of the ground state \cite{rodenburg2016}.
Working under these conditions, we obtain a MFG sensitivity of the order of 1 $\mu$Tm$^{-1}$/$\sqrt{\mbox{Hz}}$. However, the MFG sensitivity of the proposed magnetometer degrades under less stringent conditions. Specifically, we identify a suitable parameter regime where MFG sensitivity can attain a value of 100 $m$Tm$^{-1}$/$\sqrt{\mbox{Hz}}$ at room temperature and atmospheric pressure.

On the other hand, in order to manipulate the spin of the NV-center we consider a Ramsey pulse sequence \cite{rondin2014}. After initial preparation of the spin state, the spin-mechanical interaction is switched on during the free evolution time, which leads the mechanical states to acquire different phases conditioned on the spin states \cite{scala2013,wan2016}. Finally, after a suitably chosen interaction time, the acquired phase can be read out optically. Following this procedure, we obtain the photon-shot noise and spin-projection noise limited MFG sensitivity of the order of 100 $\mu$Tm$^{-1}$/$\sqrt{\mbox{Hz}}$. For these conditions, we find that the proposed magnetometer does not suffer appreciably from decoherence under realistic experimental conditions, unlike clamped oscillators \cite{forstner2012,forstner2014,yu2016}. Further, cavity optomechanical magnetometers utilize a magnetostrictive material whose geometry is difficult to control causing an undesirably weak coupling between mechanical motion and applied force \cite{forstner2012}. Contrary to this, the proposed magnetometer employs a nanodiamond with a single NV-center which are attractive systems for magnetometry \cite{maze2008}. Thus, the ability to sensitively detect magnetic field gradients, portability, small size and suitability for working under both feedback-cooled as well as room temperature  conditions, makes the proposed magnetometer appropriate for sensitive applications \cite{budker2007}. 

\section{Model}

\begin{figure}[!ht]
\begin{center}
\includegraphics[scale=0.1]{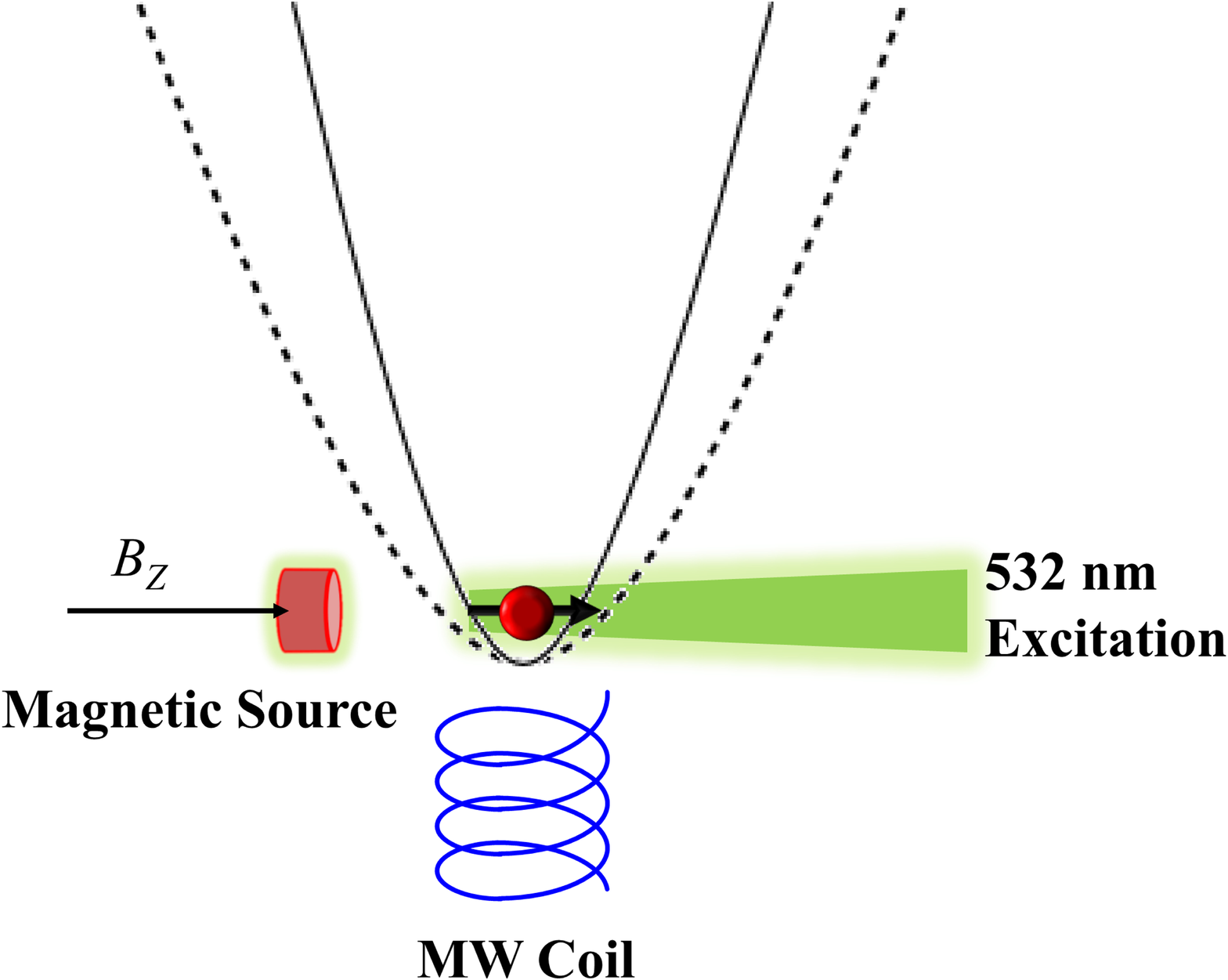}
\caption{Schematic of the optically levitated nanodiamond oscillating in dipole trap along z-direction. A green laser of wavelength 532 nm excites the nanodiamond. A MFG along z-direction is applied to engineer spin-mechanical coupling. Microwaves are used to manipulate the ground spin state of nanodiamond.}
\label{fig1}
\end{center}
\end{figure}

Recently, nanodiamonds containing single nitrogen-vacancy (NV) center have been optically levitated in vacuum \cite{levi2015}. We consider such a single NV-center diamond of mass $m$ which  is trapped in vacuum by a focused Gaussian beam resulting in mechanical oscillations at a frequency $\omega_{m}$, as shown in Fig. 1. For small amplitudes, as considered in this paper, the oscillations along the three spatial directions are uncoupled and may be considered independently. We will consider nanodiamond oscillations in the optical trap along the $z$-axis. The position of the levitated particle is continuously monitored by means of interferometric techniques. The detected position information is fed back to the trapping beam in order to modulate its optical intensity \cite{levi2015}. Such feedback causes additional damping of the nanodiamond and gives rise to cooling,  in addition to also causing some backaction heating. Under optimal conditions, the ground state of the oscillator can be prepared \cite{rodenburg2016}. Neglecting the spin degree of freedom for the moment, the quantum dynamics of the nanodiamond oscillation are modeled by the master equation \cite{rodenburg2016}

\begin{align}\label{eq1}
\dot{\rho}&=\frac{1}{i\hbar}\left[H_{m},\rho\right]-\left(\frac{A_{t}+A_{p}+D_{p}}{2}\right)\mathcal{D}\left[Q_{z}\right]\rho\left(t\right)-\frac{D_{q}}{2}\mathcal{D}\left[P_{z}\right]\rho\left(t\right)\nonumber\\
&-i\frac{\eta_{f}}{4m}\left[Q_{z},\left\{P_{z},\rho\right\}\right]+\mathcal{F}\left[\rho\left(t\right)\right]\;,
\end{align}
where, the first term represents the unitary evolution of the motion of center-of-mass mode $a_{m}$ described by the Hamiltonian $H_{m}=\hbar\omega_{m}a_{m}^{\dagger}a_{m}$. The second (third) term describes the momentum (position) diffusion of the nanodiamond due to collisions with background gas, with momentum (position) diffusion coefficient $D_{p}=2\eta_{f}k_{B}Tz_{0}^{2}/\hbar^{2} (D_{q}=\eta_{f}\hbar^{2}/24k_{B}Tm^{2}z_{0}^{2})$, where, $T$ is the gas temperature, $k_{B}$ is Boltzmann's constant and $z_{0}=\sqrt{\hbar/2m\omega_{m}}$ is the zero point fluctuation of the mechanical oscillator. Also, $A_{t}$ ($A_{p}$) represents the heating rate due to trap (probe) beam photon scattering, respectively.  The fourth term accounts for gas damping with coefficient of friction $\eta_{f}=6\pi\mu R$, where $\mu$ is the dynamic viscosity of the surrounding gas and $R$ is the radius of nanodiamond. The superoperator in the fifth term corresponds to feedback and contains both the respective feedback damping and backaction and is given by $\mathcal{F}\left[\rho\left(t\right)\right]=-i\chi^{2}\Phi G\left[Q_{z}^{3},\left\{P_{z},\rho\right\}\right]-\chi^{2}\Phi  G^{2}\mathcal{D}\left[Q^{3}_{z}\right]\rho$. Here, $\Phi$ is the average detected flux of photons, $G$ is the feedback
gain and $\chi$ is the scaled optomechanical coupling. The respective dimensionless position and momentum operators are defined as $Q_{z}=a_{m}^{\dagger}+a_{m}$ and $P_{z}=i(a_{m}^{\dagger}-a_{m})$. The Lindblad superoperator in Eq. (\ref{eq1}) is written as $\mathcal{D}\left[Q_{z}\right]\rho=Q^{\dagger}_{z}Q_{z}\rho+\rho Q^{\dagger}_{z}Q_{z}-2Q_{z}\rho Q^{\dagger}_{z}$ and the creation (annihilation) operator of the mechanical oscillator along z-axis is $a_{m}^{\dagger}$ $\left(a_{m}\right)$. In earlier works, this master equation was shown to analytically model the phonon dynamics \cite{rodenburg2016} and oscillator phase space distributions \cite{wencho2016} observed experimentally.

\begin{figure}[!ht]
\begin{center}
\begin{tabular}{cc}
\subfigure[]{\includegraphics[scale=0.2]{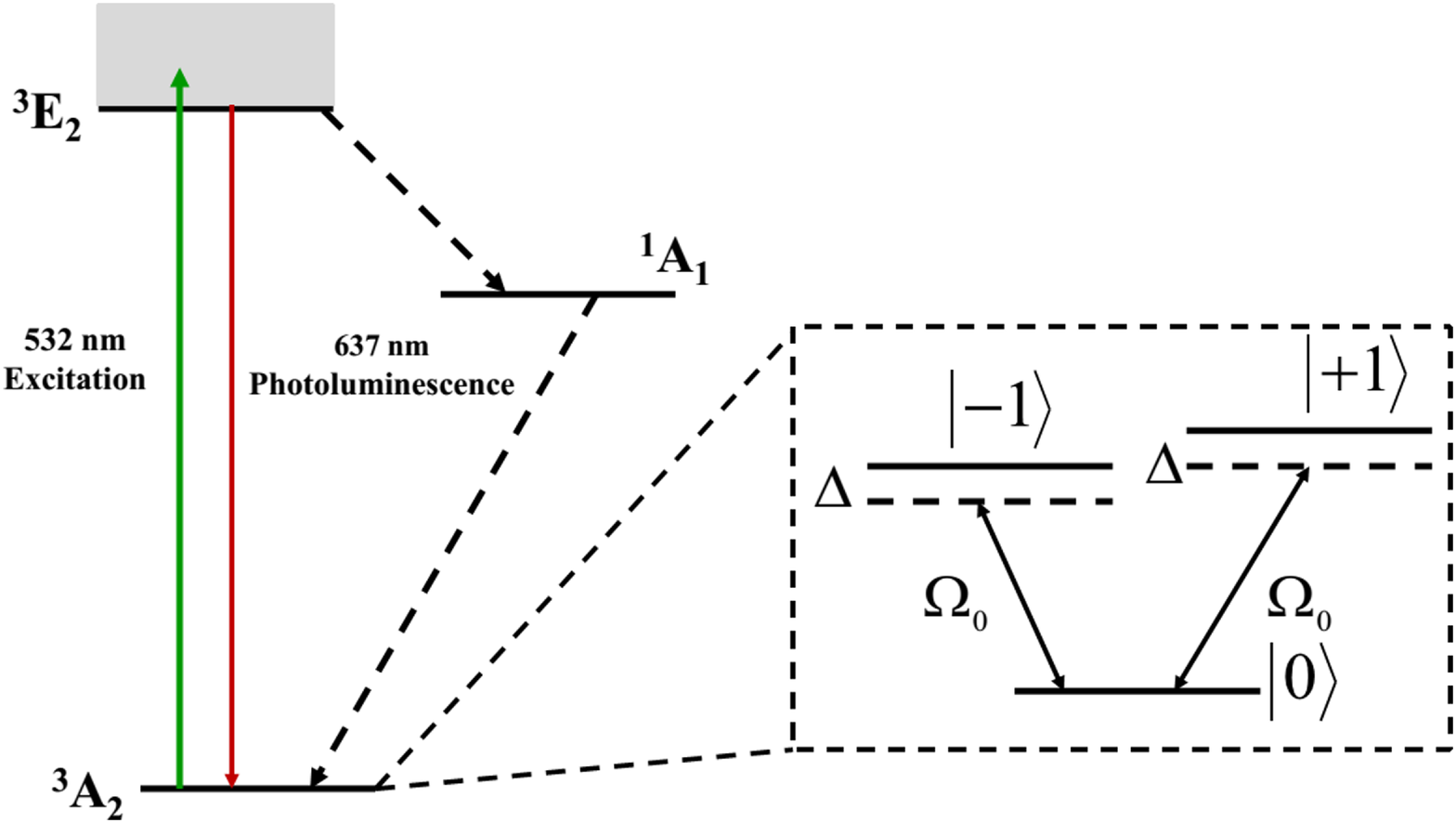}} & \subfigure[]{\includegraphics[scale=0.14]{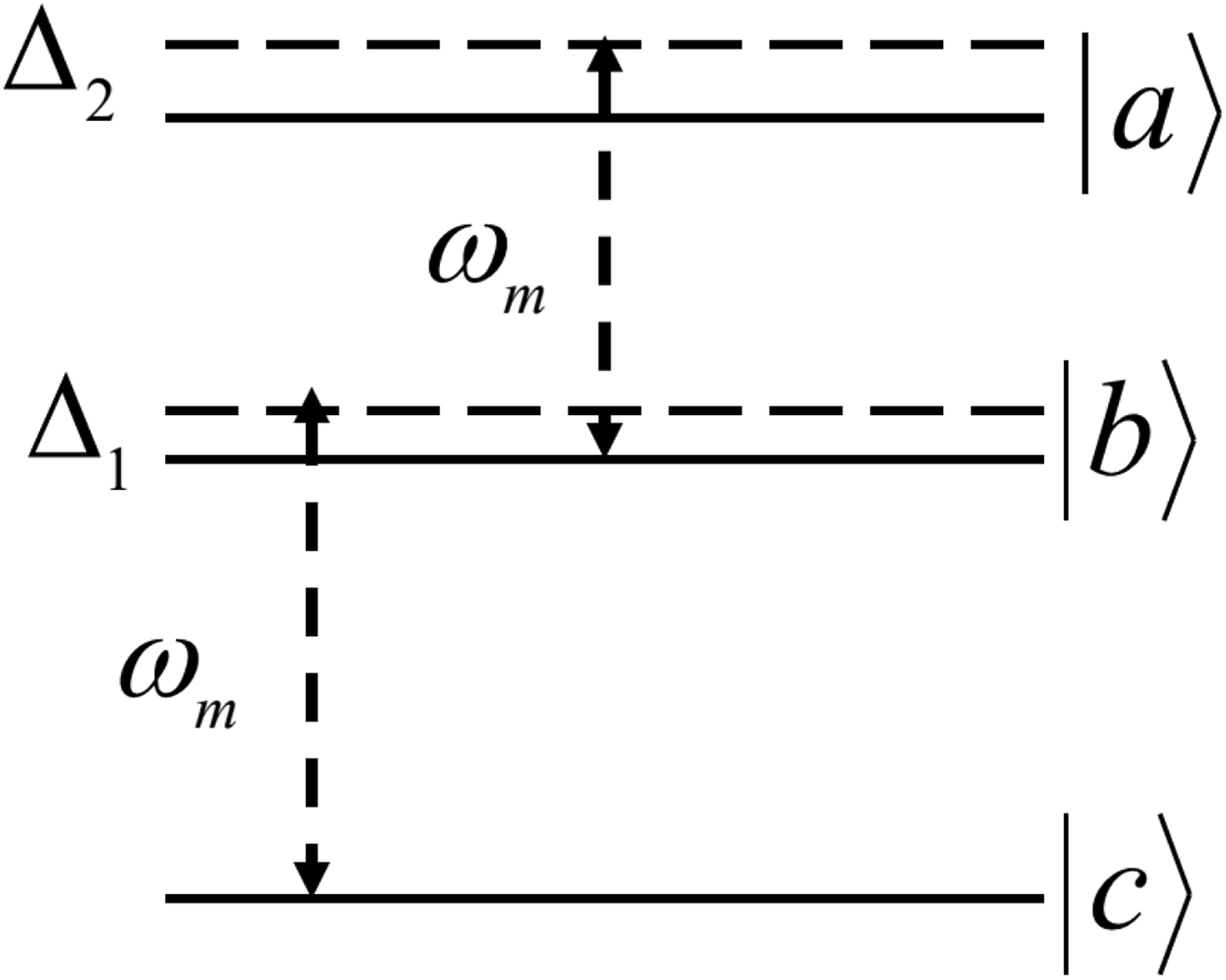}}
\end{tabular}
\caption{(a) Level diagram for excitation and read-out of NV-spins by optical means. Inset shows the bare energy-level diagram for the triplet ground-state of NV-center in nanodiamond. The microwave fields of Rabi frequency $\Omega_{0}$ drives the transitions $|0\rangle\rightarrow|+1\rangle$ and $|0\rangle\rightarrow|-1\rangle$ with detuning $\Delta$. (b) The dressed-state description of NV-spin. Phonons of energy $\omega_{m}$ drives the transitions $|c\rangle\rightarrow|b\rangle$ and $|b\rangle\rightarrow|a\rangle$ with effective detunings $\Delta_{1}=\omega_{m}-\omega_{bc}$ and $\Delta_{2}=\omega_{m}-\omega_{ab}$, respectively.}
\label{fig2}
\end{center}
\end{figure}

Besides the mechanical motion, levitated optomechanics of the nanodiamond provides a versatile platform for coherent control over spin states, as well \cite{pettit2017}. We now consider the addition of the spin degree of freedom to the dynamics. The NV defect is typically available in two different forms, namely the neutral state NV$^{0}$ and negatively charged state NV$^{-}$ \cite{rondin2014}. These two forms have different spin and optical properties. However, the negatively charged state of the defect is of particular interest for magnetometry applications, due to the existence of its spin triplet ground state  which can be initialized, coherently manipulated and read out by optical means. We consider illumination by green laser of 532 nm wavelength driving the spin-conserving transitions $^{3}A_{2}\rightarrow^{3}E_{2}$ followed by spin-selective intersystem crossing towards an intermediate spin singlet state $^{1}A_{1}$ [see Fig. 2(a)]. This leads to a high degree of spin polarization in the $m_{s}=0$ state (here, $m_{s}$ denotes the spin projection along the intrinsic quantization axis), which results in higher photoluminescence intensity from the 637 nm zero-phonon line, as shown in Fig. 2(a).

The spin triplet electronic ground state $^{3}$A$_{2}$ of the NV-center with $m_{s}=0$ and $m_{s}=\pm  1$ is shown in the inset of Fig. 2(a). The bare state Hamiltonian describing the NV spin is written as $H_{NV}=-\hbar\Delta\left(|+1\rangle\langle+1|+|-1\rangle\langle-1|\right)$, where $\Delta$ is the detuning. Now, once initially prepared in the $m_{s}=0$ state by optical means, the NV center is subjected to two microwave fields which drive the transitions $|0\rangle\leftrightarrow|+1\rangle$ and $|0\rangle\leftrightarrow|-1\rangle$ with both Rabi frequencies equal to $\Omega_{0}$. The corresponding Hamiltonian in the rotating-wave frame of the microwave field is written as $H_{drive}=\frac{\hbar \Omega_{0}}{2}\left(|0\rangle\langle +1|+|0\rangle\langle -1|+h.c.\right)$. We now apply a MFG $B_{0}=\frac{\partial B_{z}\left(z\right)}{\partial z}$ along the $z$-direction.
This creates a magnetic field $\mathbf{B}=B_{0}\vec{z}$ and generates the coupling between spin and mechanical degrees of freedom of the oscillator, governed by the interaction Hamiltonian $H_{int}=\hbar gS_{z}\left(a_{m}^{\dagger}+a_{m}\right)$, where, $S_{z}=|+1\rangle\langle +1|-|-1\rangle\langle -1|$ is the spin operator aligned along NV symmetry axis, $g=\frac{g_{l}\mu_{B}B_{0}z_{0}}{\hbar}$ is coupling strength,
$g_{l}\approx$2 is the Land\'{e} factor and $\mu_{B}$ is the Bohr magneton. 

It is to be noted that besides the center of mass motion, the torsional vibrations of levitated nanodiamonds have also been realized \cite{thai2016,delord2017}. Further, the quantum ground state cooling of torsional mode \cite{thai2016,tdelord2017} and torsional matter-wave interferometry \cite{ma2016} is proposed as well.  In these works the torsional modes can be coupled to NV center by employing a \textit{homogeneous} magnetic field and this torsional-spin coupling is proportional to the coordinate of torsional displacement.  However, in the present system we work in a regime where homogeneous magnetic field vanishes and only MFG is present, which implies to the lowest order spin-torsional coupling vanishes.  Nevertheless, MFG creates a magnetic field which is proportional to the linear vibrational mechanical amplitude. Hence the torsional-spin coupling is proportional to the product of linear times torsional mechanical amplitude which comes out to be very small in the context of present system. Typically, the particle is energetic enough to climb out of the torsional well and rotate freely, the rotational frequency is about an order of magnitude higher than the oscillation frequency and the spin-torsional interaction averages to zero \cite{arita2013}.   

Now, to accurately describe the spin-mechanical interaction, we consider the resonant interaction of microwave fields with the NV center in the dressed state eigenbasis of $H_{NV}+H_{drive}$ [see Fig. 2(b)]. These dressed states with their corresponding eigenvalues can be written as
\begin{align}\label{eq2}
|a\rangle &=\sin(\theta)|0\rangle+\cos(\theta)|+\rangle;&\omega_{a}&=\left(-\Delta+\sqrt{\Delta^{2}+2\Omega_{0}^{2}}\right)\bigg/2\nonumber\;,\\
|b\rangle&=|-\rangle;&\omega_{b}&=-\Delta\nonumber\;,\\
|c\rangle&=\cos(\theta)|0\rangle-\sin(\theta)|+\rangle;&\omega_{c}&=\left(-\Delta-\sqrt{\Delta^{2}+2\Omega_{0}^{2}}\right)\bigg/2\;,
\end{align} 
where, $|\pm\rangle=\frac{\left(|+1\rangle\pm|-1\rangle\right)}{\sqrt{2}}$ and $\tan\left(2\theta\right)=-\frac{\sqrt{2}\Omega_{0}}{\Delta}$. 
\section{Manipulation of mechanical motion}
In this section, to analyze the MFG sensitivity, we consider the manipulation of the mechanical degrees of freedom and taking trace over the spin degree of freedom. The physical idea is to investigate the effect of weak coupling of the mechanics to the spin degree of freedom which is the one addressed by the MFG. Specifically, we will analyze the position spectrum of the mechanical oscillator. As explained above, optical illumination leads to electric dipole transitions between the ground and excited states, subsequently followed by fast dissipation. This dissipation controls the populations and coherences for the NV dressed states. It is to be kept in mind that such a quick dissipation mechanism is necessary for the generation of steady state squeezing and dissipation-induced cooling leading to the preparation of ground state of the mechanical oscillator. Taking this into account, we trace over the spin degrees of freedom and include the contribution from various dissipative terms to obtain the following full master equation for the mechanical oscillator \cite{ge2016,rodenburg2016}
\begin{align}\label{eq3}
\dot{\rho}&=\frac{1}{i\hbar}\left[H_{m},\rho\right]-\left(\frac{A_{t}+A_{p}+D_{p}}{2}\right)\mathcal{D}\left[Q_{z}\right]\rho\left(t\right)-\frac{D_{q}}{2}\mathcal{D}\left[P_{z}\right]\rho\left(t\right)-i\frac{\eta_{f}}{4m}\left[Q_{z},\left\{P_{z},\rho\right\}\right]\nonumber\\
&+\mathcal{F}\left[\rho\left(t\right)\right]-\frac{\mathcal{A}_{-}}{2}\mathcal{D}\left[a_{m}\right]\rho-\frac{\mathcal{A}_{+}}{2}\mathcal{D}\left[a_{m}^{\dagger}\right]\rho-\frac{i\delta}{2}\left[a_{m}^{\dagger}a_{m},\rho\right]\;,
\end{align}

where, $\mathcal{A}_{-}=2g^{2}\alpha_{1}$ ($\mathcal{A}_{+}=2g^{2}\alpha_{2}$) and $\delta=2g^{2}\alpha_{3}$ describe dissipation induced cooling (heating) and mechanical frequency shift due to spin mechanical coupling. Here, $\alpha_{1}, \alpha_{2}$ and $\alpha_{3}$ can be written as \cite{ge2016}
\begin{align}\label{eq4}
\alpha_{1}&=\mbox{Re}\left[\frac{\sin^{2}\left(\theta\right)}{\mathcal{N}}\left(p_{2}\rho_{cc}+p_{3}\rho_{ca}\right)+\frac{\cos^{2}\left(\theta\right)}{\mathcal{N}}p_{1}\rho_{bb}\right]\;,\nonumber\\
\alpha_{2}&=\mbox{Re}\left[\frac{\cos^{2}\left(\theta\right)}{\mathcal{N}}\left(p_{1}\rho_{aa}+p_{3}\rho_{ac}\right)+\frac{\sin^{2}\left(\theta\right)}{\mathcal{N}}p_{2}\rho_{bb}\right]\;,\\
\alpha_{3} &=\mbox{Im}\left[\frac{\sin^{2}\left(\theta\right)\left(p_{2}\left(\rho_{cc}-\rho_{bb}\right)+p_{3}\rho_{ca}\right)}{\mathcal{N}}+\frac{\cos^{2}\left(\theta\right)\left(p_{1}\left(\rho_{aa}-\rho_{bb}\right)+p_{3}\rho_{ac}\right)}{\mathcal{N}}\right]\;,\nonumber
\end{align}
where, $p_{1}=-i\Delta_{1}+\frac{\Gamma_{1}}{2}\left(1+\sin^{2}\theta\right)$, $p_{2}=i\Delta_{2}+\frac{\Gamma_{1}}{2}\left(1+\cos^{2}\theta\right)$, $p_{3}=\frac{\Gamma_{1}}{4}\sin\left(2\theta\right)$, $\mathcal{N}=p_{1}p_{2}-p_{3}^{2}$ and $\Gamma_{1}\approx\Gamma_{2}$ is the decay of states $|+1\rangle$ and $|-1\rangle$, respectively. The detailed derivation of population and coherence terms in Eq. (\ref{eq4}) is presented in \cite{ge2016}. 
 
\subsection{Position spectrum}
Now to continuously monitor the position of nanodiamond, the master equation Eq. (\ref{eq3}) can be unraveled in terms of the following set of quantum Langevin equations for position and momentum quadratures \cite{rodenburg2016}
\begin{align}
\dot{Q}_{z}&=\left(\omega_{m}+\frac{\delta}{2}\right)P_{z}-\left(\frac{\mathcal{A}_{-}-\mathcal{A}_{+}}{2}\right)Q_{z}\label{eq5}\;,\\
\dot{P}_{z}&=-\left(\omega_{m}+\frac{\delta}{2}\right)Q_{z}-\left[ \Gamma+\left(\frac{\mathcal{A}_{-}-\mathcal{A}_{+}}{2}\right) \right] P_{z}\nonumber\\
&+\sqrt{2m\Gamma_{0}k_{B}T_{eff}}\xi_{T}\left(t\right)+\sqrt{18\eta\Phi_{n}G^{2}Q^{4}}\xi_{F}\left( t\right)+\left[\sqrt{2\mathcal{A}_{+}}+\langle N\rangle\left(\sqrt{2\mathcal{A}_{+}}-\sqrt{2\mathcal{A}_{-}}\right)\right]\xi_{S}\left( t\right)\;,\label{eq6}
\end{align}
where, $\xi\left(t\right)$ are white noise terms with zero mean and correlation $\mbox{E}\left[\xi(t)\xi(t^{\prime})\right]=\delta\left(t-t^{\prime}\right)$, and $T_{eff}$ is the effective temperature of the total background due to the combination of gas and optical scattering. The parameter $\Gamma=\Gamma_{0}+\delta\Gamma$, where $\Gamma_{0}$ represents the gas damping and $\delta\Gamma\approx 12\chi^{2}G\left(\langle N\rangle+\frac{1}{2}\right)$. Note that last stochastic term in Eq. (\ref{eq6}) arises as a consequence of the spin-mechanical coupling and depends on the mechanical state of the system. Using Eqs. (\ref{eq5},\ref{eq6}), the stochastic differential equation for position ($q_{z}=z_{0}Q_{z}$) can be written as
\begin{align}
&\ddot{q}_{z}+\left[\Gamma+\mathcal{A}_{-}-\mathcal{A}_{+}\right]\dot{q}_{z}+\left\{\left(\omega_{m}+\frac{\delta}{2}\right)^{2}+\left[\Gamma+\left(\frac{\mathcal{A}_{-}-\mathcal{A}_{+}}{2}\right)\right]\left(\frac{\mathcal{A}_{-}-\mathcal{A}_{+}}{2}\right)\right\}q_{z}\nonumber\\
&=\frac{\left(F_{T}+F_{F}+F_{S}\right)}{m}\label{eq7}\;,
\end{align}
where,   $F_{T}, F_{F}$ and $F_{s}$ are independent stochastic forces due to thermal, feedback backaction heating and spin interaction, respectively with zero mean and correlations $\langle F_{T}\left(t\right)F_{T}\left(t^{\prime}\right)\rangle=\mathcal{S}_{T}\delta\left(t-t^{\prime}\right)$, $\langle F_{F}\left(t\right)F_{F}\left(t^{\prime}\right)\rangle=\mathcal{S}_{F}\delta\left(t-t^{\prime}\right)$ and $\langle F_{S}\left(t\right)F_{S}\left(t^{\prime}\right)\rangle=\mathcal{S}_{S}\delta\left(t-t^{\prime}\right)$. Here, 
\begin{align}
\mathcal{S}_{T}&=\left(1+\frac{\delta}{2\omega_{m}}\right)2m\Gamma_{0}k_{B}T_{eff}\label{eq8}\;,\\
\mathcal{S}_{F}&=\left(1+\frac{\delta}{2\omega_{m}}\right)54m\hbar\omega_{m}\chi^{2}\Phi G^{2}\left(2\langle N\rangle^{2}+2\langle N\rangle+1\right)\label{eq9}\;,\\ 
\mathcal{S}_{S}&=\left(1+\frac{\delta}{2\omega_{m}}\right)m\hbar\omega_{m}\left[\sqrt{\mathcal{A}_{+}}+\langle N\rangle\left(\sqrt{\mathcal{A}_{+}}-\sqrt{\mathcal{A}_{-}}\right)\right]\label{eq10}\;.
\end{align}
The position spectrum of the mechanical oscillator can be obtained by taking the Fourier Transform of Eq. (\ref{eq7}) which gives  
\begin{align}\label{eq11}
\tilde{q}_{z}\left(\omega\right)&=\chi_{m}\left(\omega\right)\left[\tilde{F}_{T}+\tilde{F}_{F}+\tilde{F}_{S}\right]\;,
\end{align}
where, $\chi_{m}\left(\omega\right)$ represents the optomechanical susceptibility of the oscillator and  is given by
\begin{align}\label{eq12}
\chi_{m}\left(\omega\right)=\frac{1}{m\Bigg\{\left[\left(\omega_{m}+\frac{\delta}{2}\right)^{2}+\left(\frac{\mathcal{A}_{-}-\mathcal{A}_{+}}{2}\right)\left(\Gamma+\frac{\mathcal{A}_{-}-\mathcal{A}_{+}}{2}\right)-\omega^{2}\right]-i\left(\Gamma+\mathcal{A}_{-}-\mathcal{A}_{+}\right)\omega\Bigg\}}\;.
\end{align}
The expectation value of $\tilde{q}_{z}$ is therefore $\langle \tilde{q}_{z}\left(\omega\right)\rangle=0$ and the positional power spectral density (PSD) is written as
\begin{align}\label{eq13}
\langle|\delta\tilde{q}_{z}\left(\omega\right)|^{2}\rangle=\vert\chi_{m}\left(\omega\right)\vert^{2}\left[\langle|\tilde{F}_{T}|^{2}\rangle+\langle|\tilde{F}_{F}|^{2}\rangle+\langle|\tilde{F}_{S}|^{2}\rangle\right]\;.
\end{align}
Thus, the position noise spectrum including the contribution from shot-noise is given by
\begin{align}\label{eq14}
\langle|\delta\tilde{q}_{z}\left(\omega\right)|^{2}\rangle&=\vert\chi_{m}\left(\omega\right)\vert^{2}\left[\mathcal{S}_{T}+\mathcal{S}_{F}+\mathcal{S}_{S}\right]+\frac{z_{0}^{2}}{\chi^{2}\Phi}\;.
\end{align}
The position noise spectrum in Eq. (\ref{eq14}) depends upon the spin-mechanical coupling and hence on the MFG. We explore this dependence of the noise spectrum to describe the sensitivity of MFG. To do so, we work in a regime of low pressure where cooling from various dissipative mechanisms dominates over heating. For instance, the \textit{optimal feedback} cooling at pressure $\lesssim$ 10$^{-5}$ mbar assists to prepare the ground state of the mechanical oscillator with $\langle N\rangle <1$ \cite{rodenburg2016}. For the spin part, we consider the cooling transition $\left(|c\rangle\rightarrow|b\rangle\right)$ to be resonant $\left(\Delta_{1}=0\right)$ while the heating transition $\left(|a\rangle\rightarrow|c\rangle\right)$ is assumed to be far off-resonant $\left(\Delta_{2}=-\Delta\right)$. In this situation, $\mathcal{A}_{+}=0$, $\delta=0$ and $\mathcal{A}_{-}$ simplifies to first order as \cite{ge2016} $\mathcal{A}_{-}\approx g^{2}\alpha$, where 
\begin{align}\label{eq15}
\alpha=\frac{4}{\Gamma_{1}}\frac{\left(1+\cos\left(2\theta\right)\right)\sin^{2}\left(2\theta\right)}{9-\cos^{2}\left(2\theta\right)}\;.
\end{align}
The cooling term as described above varies with microwave power. It is possible to cool the mechanical oscillator with $\langle N\rangle < 1$ by working in a suitable regime of microwave power. For instance, at $\Omega_{0}=0.8\omega_{m}$ dissipative cooling dominates over heating and the steady state mean phonon number approximates to 0.3 \cite{ge2016}.

\begin{figure}[!ht]
\begin{center}
\begin{tabular}{cc}
\includegraphics[scale=0.45]{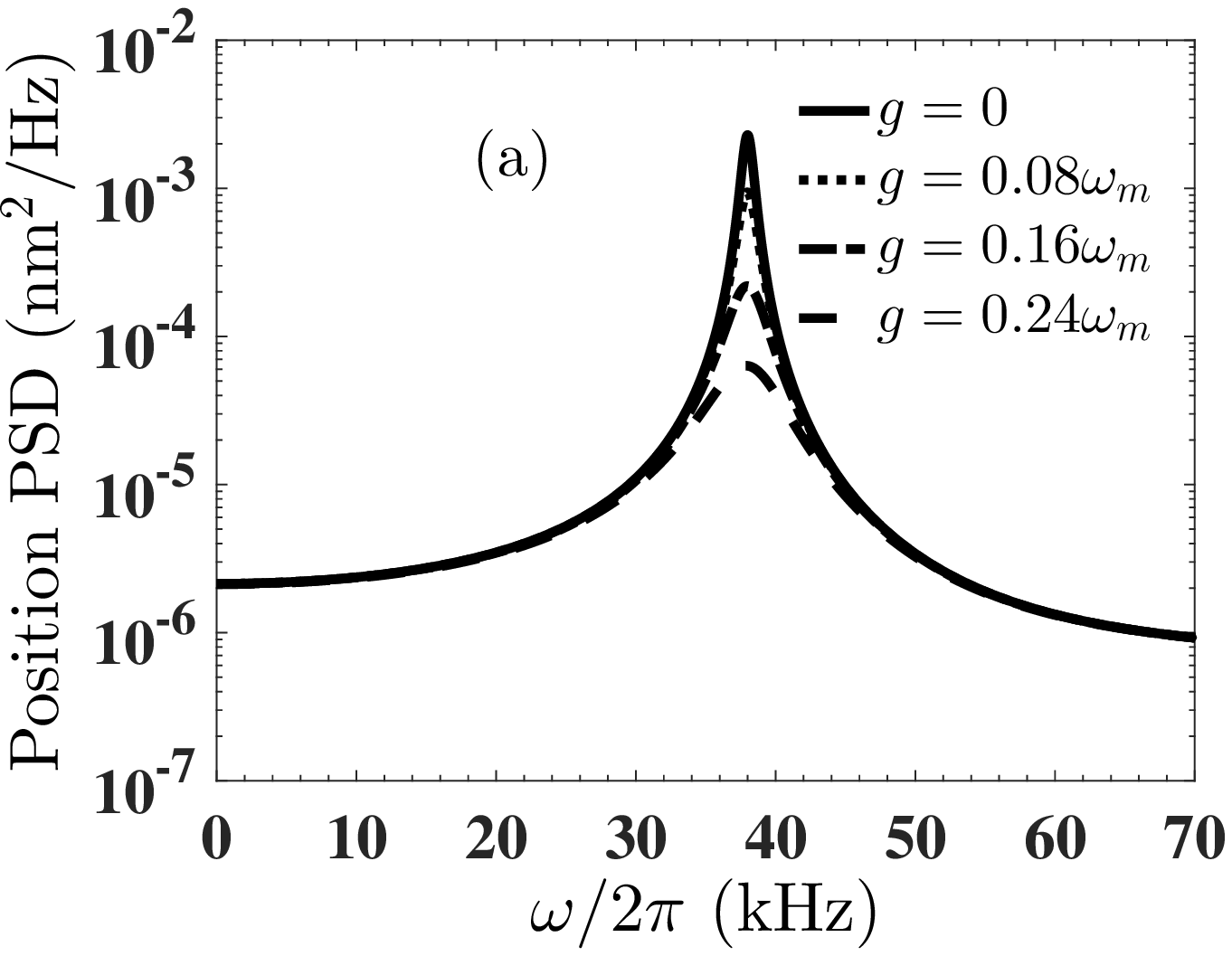} & \includegraphics[scale=0.45]{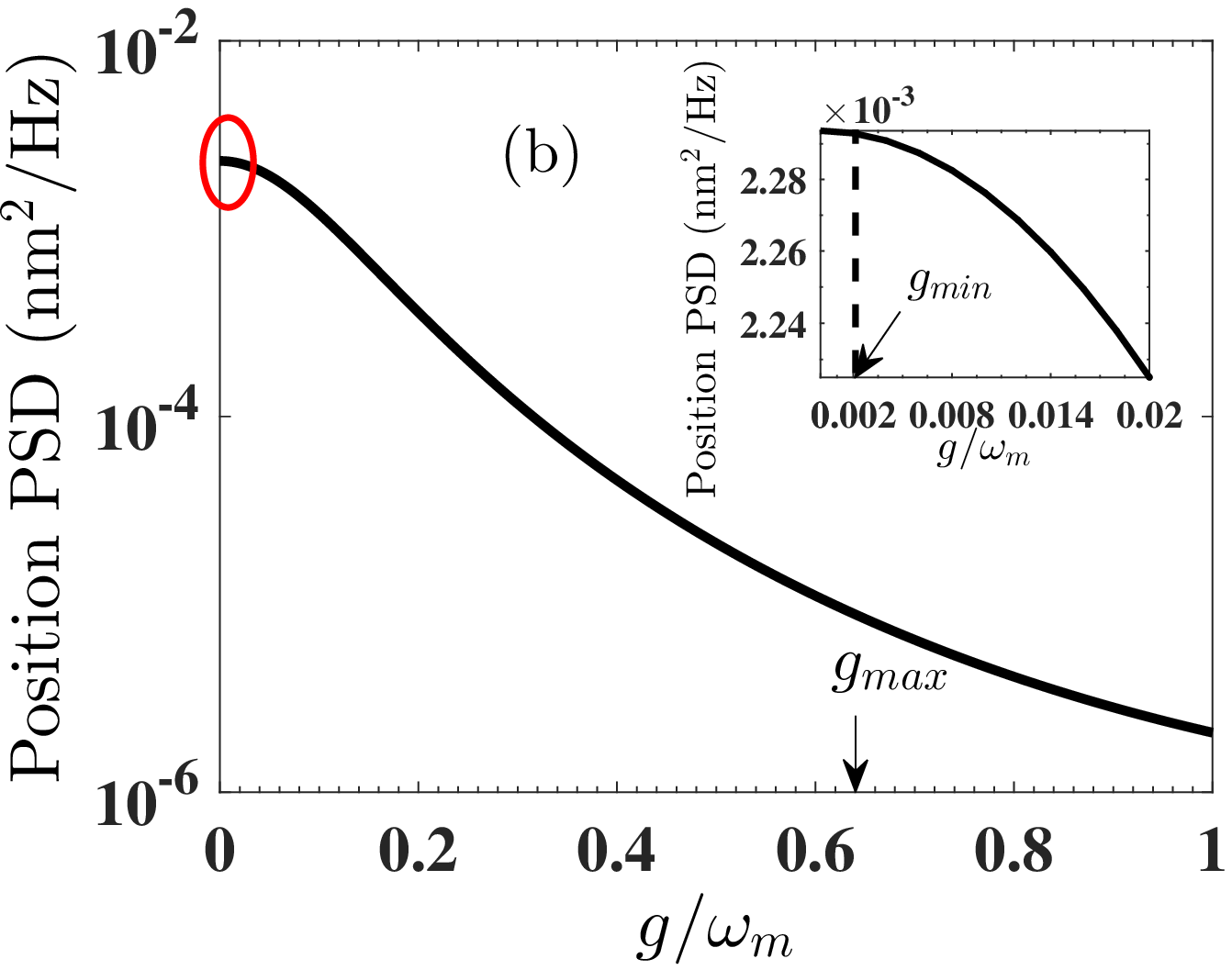}
\end{tabular}
\caption{(a) Position power spectral density for z-component of motion versus frequency. The parameters chosen are $\omega_{m}=38 \mbox{kHz}$, $R=$50 nm, density=2200 kg/m$^{3}$, $T_{eff}=4$ K, $\chi=10^{-7}$, $G=1/18$, $\Omega_{0}=0.8\omega_{m}$, $\Gamma_{1}=0.25\omega_{m}$. (b) Position power spectral density versus scaled spin-mechanical coupling at $\omega=\omega_{m}$. The inset shows the position PSD corresponding to red ellipse for smaller values of $g/\omega_{m}$.}
\label{fig3}
\end{center}
\end{figure}

Working under the conditions stated above, we plot the position noise spectrum from Eq. (\ref{eq14}) in Fig. \ref{fig3}. In the absence of spin-mechanical coupling ($\mathcal{S}_{S}=0$), the noise spectrum is dominated by the feedback cooling \cite{rodenburg2016} term proportional to  $\mathcal{S}_{F}$ in Eq. (\ref{eq14}). However, the amplitude of the spectrum decreases at resonance as the spin-mechanical coupling is introduced [see Fig. 3(a)]. Our analysis indicates that the noise spectrum in Fig. \ref{fig3} is dominated by $|\chi_{m}\left(\omega\right)|^{2}$ [see Eq. (\ref{eq14})], and the reduction in amplitude of position PSD at resonance is a consequence of the increase in the damping of the mechanical oscillator caused by spin-mechanical interaction \cite{vitali2002}. The  susceptibility at the slightly shifted mechanical resonance $\omega=\sqrt{\omega_{m}^{2}+\frac{\mathcal{A}_{-}}{2}\left(\Gamma+\frac{\mathcal{A}_{-}}{2}\right)-\frac{1}{2}\left(\Gamma+\mathcal{A}_{-}\right)^{2}}\approx\omega_{m}$ attains the following form
\begin{align}\label{eq16}
|\chi_{m}\left(\omega_{m}\right)|^{2}_{max}=\Bigg[m^{2}\Bigg\{\left(\Gamma+\mathcal{A}_{-}\right)^{2}\left[\left(\omega_{m}^{2}+\frac{\mathcal{A}_{-}}{2}\left(\Gamma+\frac{\mathcal{A}_{-}}{2}\right)\right)-\frac{1}{2}\left(\Gamma+\mathcal{A}_{-}\right)^{2}\right]\Bigg\}\Bigg]^{-1}\;.
\end{align} 

It follows from Eq. (\ref{eq16}) that $|\chi_{m}\left(\omega_{m}\right)|_{max}^{2}$ decreases with increase in $\mathcal{A}_{-}$ and hence with increase in the spin-mechanical coupling ($g$), [see Eq. (\ref{eq15})]. In Fig. \ref{fig3}(b), we depict the decrease in the amplitude of noise spectrum (at resonance) with spin-mechanical coupling. It follows that there is a maximum value of spin-mechanical coupling ($g_{max}$=0.62$\omega_{m}$) at which the reduction in position PSD at resonance is maximum and beyond which noise is larger than the peak value of the position measurement at resonance. This value is indicated by a dotted line in Fig. \ref{fig3}(b). We also note that the smaller values of the spin-mechanical coupling would not be useful as for these values it would be difficult to detect the change in the peak of position PSD with respect to the case when coupling is absent. The minimum coupling for which such a change can be detected corresponds to $g_{min}=2\times 10^{-3}\omega_{m}$, as shown in the inset of Fig. \ref{fig3}(b).

\subsection{Magnetic field gradient sensitivity}

In the preceding analysis, we showed that the spin-mechanical coupling created by MFG substantially affects the position noise spectrum. This position PSD can be treated as a signal for which MFG sensitivity can be written as \cite{rondin2014,xu2014}
\begin{align}\label{eq17}
\eta_{B}=\frac{\langle\vert\delta\tilde{q}_{z}\left(\omega\right)|^2\rangle}{\frac{\partial \langle\vert\delta\tilde{q}_{z}\left(\omega\right)|^2\rangle }{\partial B_{0}}}\times\sqrt{t_{m}}\;,
\end{align}
where, $t_{m}$ is the measurement time. Under cooling conditions, MFG sensitivity can be written as,
\begin{align}\label{eq18}
\eta_{B}&=\frac{\langle\vert\delta\tilde{q}_{z}\left(\omega\right)|^2\rangle}{\frac{\mu_{B}g_{l}z_{0}}{\hbar}\vert\chi_{m}\left(\omega\right)\vert^{2}\left[\mathcal{D}+m\hbar\omega_{m}\langle N\rangle\sqrt{\alpha}\right]}\;,
\end{align}
where, $\mathcal{D}$ is given by
\begin{align}\label{eq19}
\mathcal{D}=2m^{2}g\alpha\left(\Gamma+\mathcal{A}_{-}\right)\vert\chi_{m}\left(\omega\right)\vert^{2}\left(\mathcal{S}_{T}+\mathcal{S}_{F}+\mathcal{S}_{S}\right)\left[\omega_{m}^{2}+\omega^{2}+\frac{\mathcal{A}_{-}}{2}\left(\Gamma+\frac{\mathcal{A}_{-}}{2}\right)\right]\;.
\end{align}
 
\begin{figure}[ht!]
\begin{center}
\begin{tabular}{cc}
\includegraphics[scale=0.45]{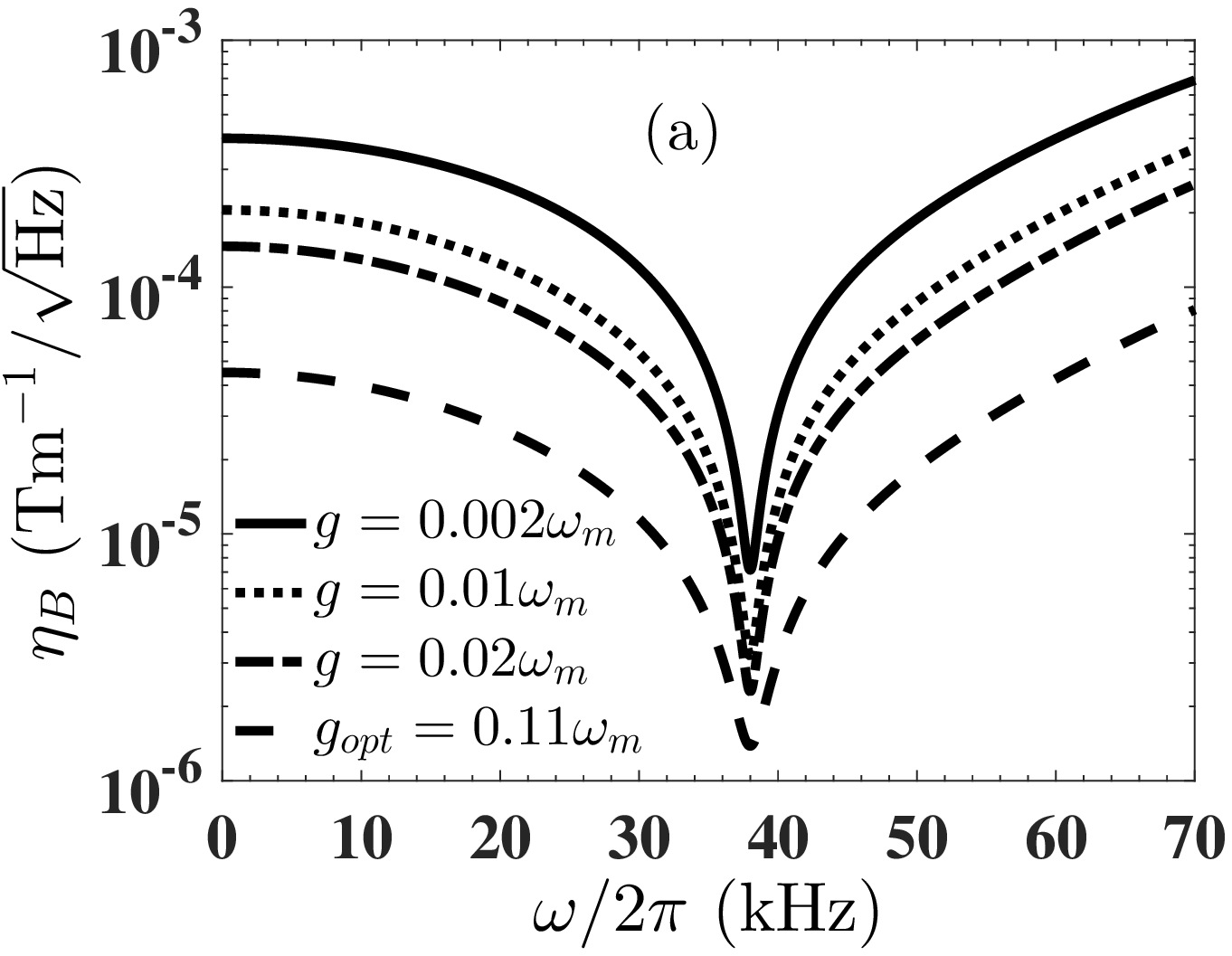} & \includegraphics[scale=0.45]{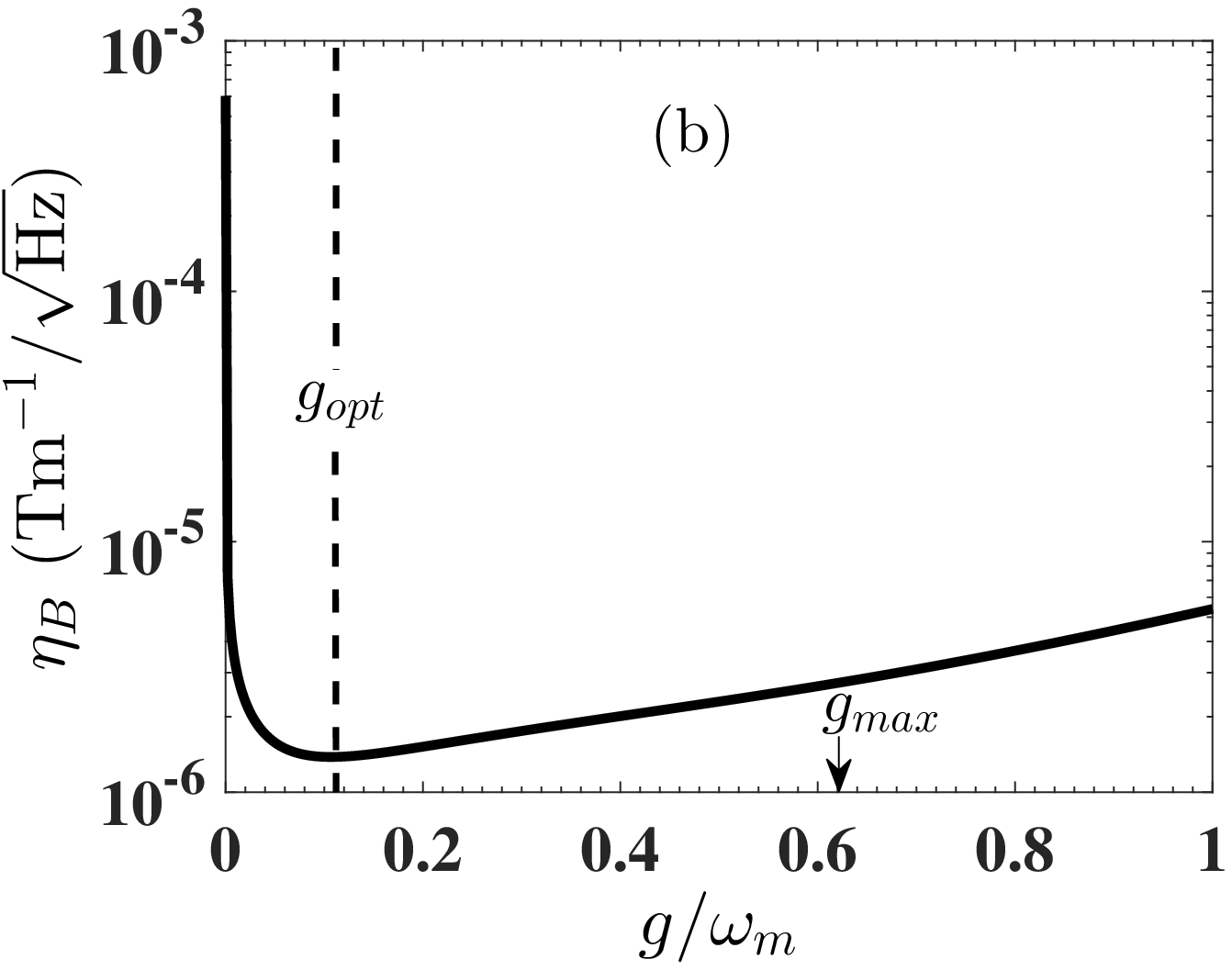}
\end{tabular}
\caption{MFG sensitivity versus (a) frequency (b) scaled spin-mechanical coupling for $t_{m}=\frac{2\pi}{\omega_{m}}$. Other parameters are same as in Fig. \ref{fig3}.}
\label{fig4}
\end{center}
\end{figure}

The sensitivity spectrum of MFG is shown in Fig. \ref{fig4}. The peak value of MFG sensitivity decreases with increase in spin-mechanical coupling and can attain a value of 1 $\mu\mbox{Tm}^{-1}/\sqrt{\mbox{Hz}}$ at $g=0.11\omega_{m}$, as shown in Fig. \ref{fig4}(a). Interestingly, there exists an optimum value of spin-mechanical coupling ($g_{opt}$=0.11$\omega_{m}$) such that for $g<g_{opt}$ MFG sensitivity improves with increase in $g$ and for $g>g_{opt}$ its value start increasing, as depicted in Fig. \ref{fig4}(b). For $g<g_{opt}$, MFG sensitivity is dominated by high spin-mechanical noise and mechanical susceptibility, both of which diminish with increase in $g$. This results in the improvement of sensitivity reaching a minimum value at $g=g_{opt}$. However, for $g>g_{opt}$, spin-mechanical noise becomes flat and susceptibility decreases causing the MFG sensitivity to increase with increase in spin-mechanical coupling. But still in this regime it can acquire a value of the order of 2 $\mu\mbox{Tm}^{-1}/\sqrt{\mbox{Hz}}$ at $g=g_{max}$.

\begin{figure}[!ht]
\begin{center}
\begin{tabular}{ccc}
&\subfigure[]{\includegraphics[scale=0.435]{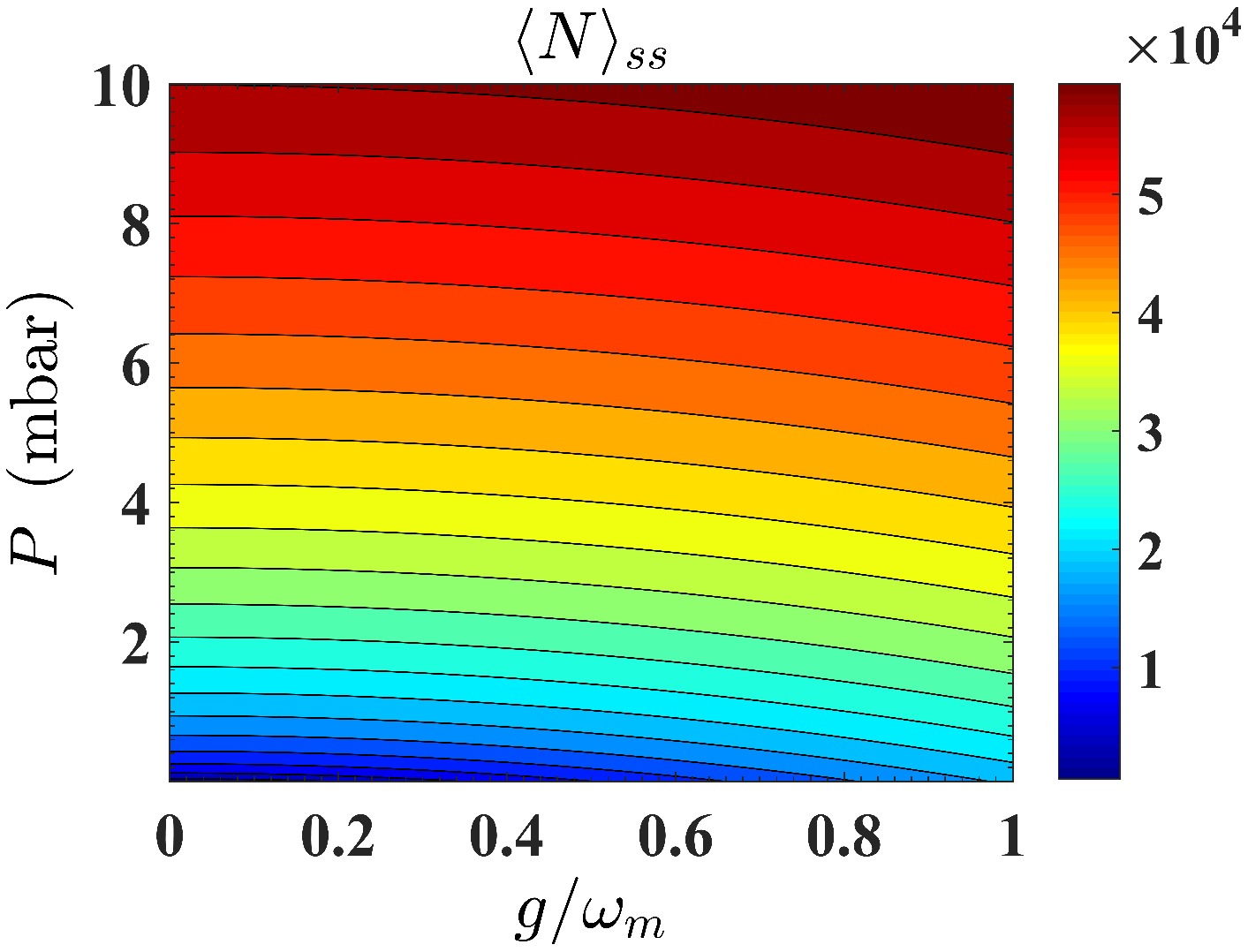}} \subfigure[]{\includegraphics[scale=0.435]{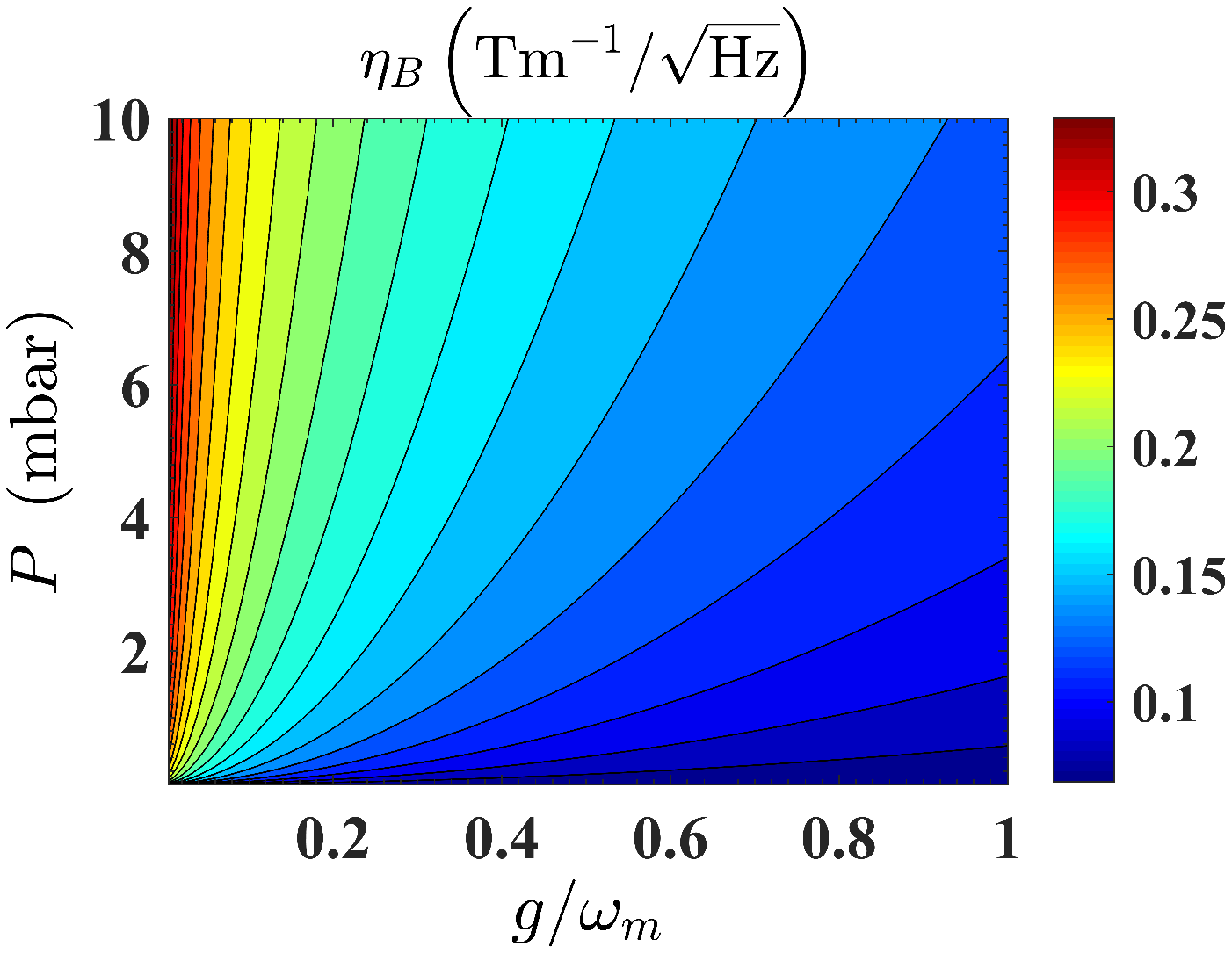}}&\\
&\subfigure[]{\includegraphics[scale=0.435]{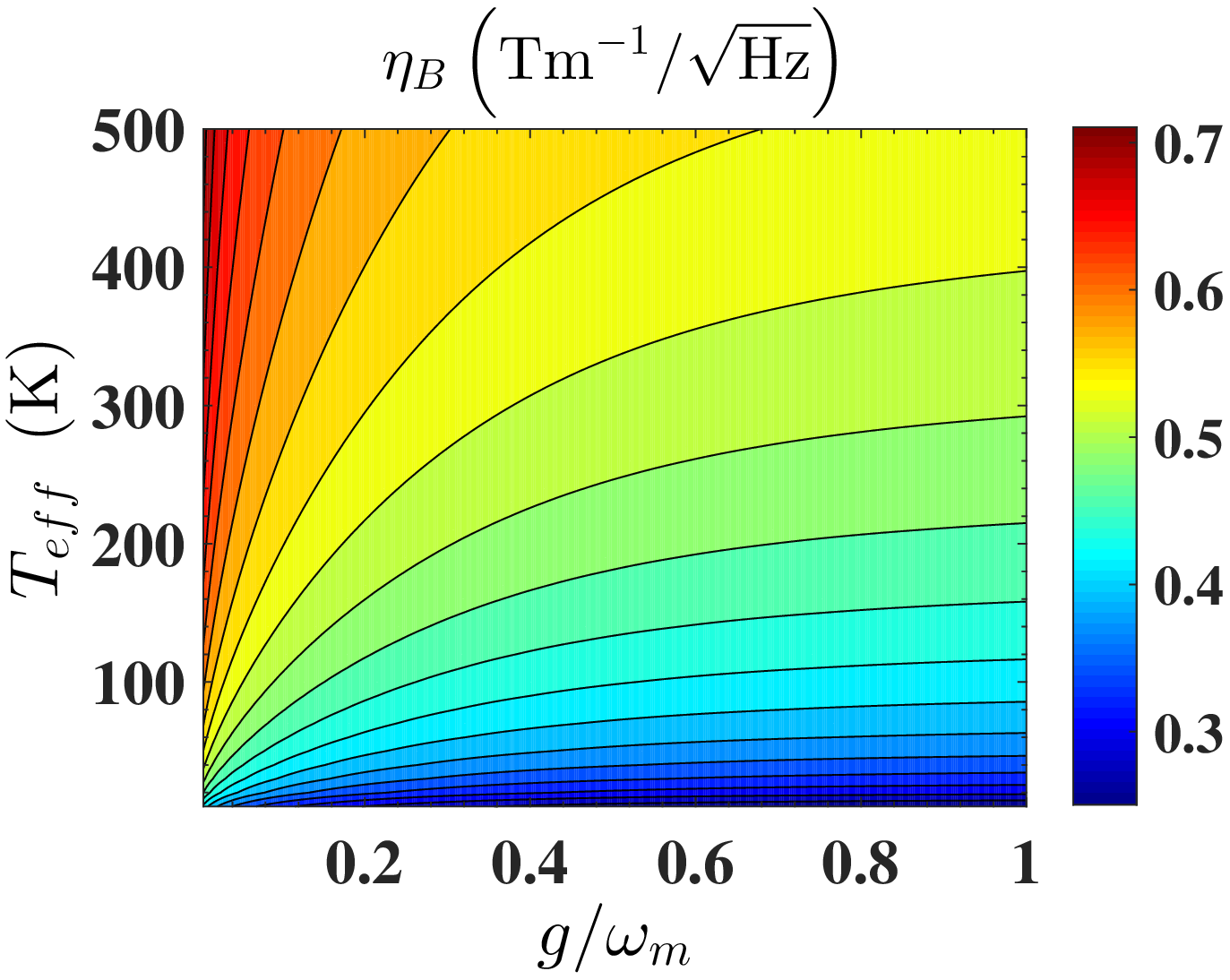}}&
\end{tabular}
\caption{(a) Steady state phonon number versus $g/\omega_{m}$ and Pressure (mbar). MFG sensitivity versus (b)$g/\omega_{m}$ and Pressure (mbar), (c) $g/\omega_{m}$ and Temperature (K). In plots (a) and (b) $T_{eff}=$300 K, $\gamma_{opt}=$1 kHz, while in (c) pressure is 0.3 mbar. The other parameters are same as in Fig. \ref{fig3}.}
\label{fig5}
\end{center}
\end{figure}

So far, the above analysis is based on the preparation of ground state for a rather low value of pressure ($\lesssim 10^{-5}$ mbar). However, to illustrate the performance of the proposed magnetometer for higher pressure and temperature, we first write the equation for the phonon number dynamics \cite{rodenburg2016} using Eq. (\ref{eq3})
\begin{align}\label{eq20}
\langle \dot{N}\rangle=-2\mathcal{J}\langle N\rangle^{2}-\left(\mathcal{J}+\mathcal{K}\right)\langle N\rangle+\mathcal{M}\;,
\end{align}
where, $\mathcal{J}=12\left[G-9G^{2}\right]\chi^{2}\Phi$, $\mathcal{K}=\eta_{f}/m+\mathcal{J}+\mathcal{A}_{-}+\mathcal{A}_{+}$ and $\mathcal{M}=D_{p}+\gamma_{opt}+\mathcal{A}_{+}$, and $\gamma_{opt}$ governs the optical scattering rate. The steady state phonon number from Eq. (\ref{eq20}) can be written as
\begin{align}\label{eq21}
\langle N\rangle_{ss}\approx\sqrt{\frac{D_{p}+\gamma_{opt}+\mathcal{A}_{+}}{2\mathcal{J}}}\;.
\end{align}
This approximation holds for $N_{0}=k_{B}T_{eff}/\hbar\omega_{m}\gg1$, where $N_{0}$ is the initial number of phonons. Note that MFG sensitivity is influenced by the state of the system. On this note, we exhibit in Fig. \ref{fig5}(a) that steady state phonon number increases with pressure for a particular value of $g$. However, the increase in spin-mechanical coupling leads to decrease in steady state phonon number for a specific value of pressure [see Fig. \ref{fig5}(a)]. 

Now, the MFG sensitivity from Eq. (\ref{eq17}) for high pressure and temperature can be written as,
\begin{align}\label{eq22}
\eta_{B}=\frac{\hbar}{\mu_{B}g_{l}z_{0}}\frac{\langle\vert\delta\tilde{q}_{z}\left(\omega\right)|^2\rangle}{\vert\chi_{m}(\omega)\vert^{2}\left(\mathcal{D}_{1}+\mathcal{D}_{2}+\mathcal{D}_{3}\right)+\mathcal{D}_{4}\left(\mathcal{S}_{T}+\mathcal{S}_{F}+\mathcal{S}_{S}\right)}\;,
\end{align}
where, $\mathcal{D}_{i}~(i=1,2,3,4)$ are given by

\begin{align}
\mathcal{D}_{1}&=\frac{2g\alpha_{3}}{\omega_{m}}2m\Gamma_{0}K_{B}T_{eff}\label{eq23}\;,\\
\mathcal{D}_{2}&=108mg\hbar\omega_{m}\chi^{2}\Phi G^{2}\left[\frac{\alpha_{3}}{\omega_{m}}\left(2\langle N\rangle^{2}+2\langle N\rangle+1\right)+\left(1+\frac{\delta}{2\omega_{m}}\right)\frac{\alpha_{2}}{\mathcal{J}}\left(2+\frac{1}{\langle N\rangle}\right)\right]\label{eq24}\;,\\
\mathcal{D}_{3}&=m\hbar\omega_{m}2g\Bigg[\frac{\alpha_{3}}{\omega_{m}}\left\{\sqrt{\mathcal{A}_{+}}+\langle N\rangle\left(\sqrt{\mathcal{A}_{+}}-\sqrt{\mathcal{A}_{-}}\right)\right\}+\left(1+\frac{\delta}{2\omega_{m}}\right)\Bigg\{\frac{\alpha_{2}}{\sqrt{\mathcal{A}_{+}}}\nonumber\\
& +\langle N\rangle\left(\frac{\alpha_{2}}{\sqrt{\mathcal{A}_{+}}}-\frac{\alpha_{1}}{\sqrt{\mathcal{A_{-}}}}\right)+\frac{1}{2\langle N\rangle}\frac{\alpha_{2}}{\mathcal{J}}\left(\sqrt{\mathcal{A}_{+}}-\sqrt{\mathcal{A}_{-}}\right)\Bigg\}\Bigg]\label{eq25}\;,\\
\mathcal{D}_{4}&=-8m^{2}g\vert\chi_{m}\left(\omega\right)\vert^{4}\Bigg\{\Bigg[\left(\omega_{m}+\frac{\delta}{2}\right)^{2}+\left(\frac{\mathcal{A}_{-}-\mathcal{A}_{+}}{2}\right)\left(\Gamma+\frac{\mathcal{A}_{-}-\mathcal{A}_{+}}{2}\right)-\omega^{2}\Bigg]\nonumber\\
&\times\Bigg[\alpha_{3}\left(\omega_{m}+\frac{\delta}{2}\right)+\left(\frac{\alpha_{1}-\alpha_{2}}{2}\right)\left(\Gamma+\mathcal{A}_{-}-\mathcal{A}_{+}\right)\Bigg]+\left(\alpha_{1}-\alpha_{2}\right)\left(\Gamma+\mathcal{A}_{-}-\mathcal{A}_{+}\right)\omega^{2}\Bigg\}\label{eq26}\;.
\end{align}

Using Eq. (\ref{eq22}), the influence of pressure and temperature on MFG sensitivity is shown in Fig. \ref{fig5}. For small spin-mechanical coupling the increase in pressure degrades the MFG sensitivity, as depicted in Fig. \ref{fig5}(b). However, for a particular value of pressure, MFG sensitivity still improves with spin-mechanical coupling for the same reasons as explained after Eq. (\ref{eq19}). For instance at a particular pressure, for a regime of $g\geq0.4\omega_{m}$, the MFG sensitivity improves and can even attain a value of the order of 100 $m$Tm$^{-1}$/$\sqrt{\mbox{Hz}}$, as shown in Fig. \ref{fig5}(b). Further, the MFG sensitivity is influenced by temperature as exhibited in Fig. \ref{fig5}(c). Again the rise in temperature degrades the MFG sensitivity. However, for a particular temperature, a strong spin-mechanical coupling improves the sensitivity and at room temperature, the improvement is of the order of 500 $m$Tm$^{-1}$/$\sqrt{\mbox{Hz}}$ for increase in spin-mechanical coupling [see Fig. \ref{fig5}(c)]. Thus, the present hybrid nanomechanical sensor harnesses useful MFG sensitivity even under room temperature and pressure conditions (albeit in the range of $m\mbox{Tm}^{-1}/\sqrt{\mbox{Hz}}$).

\section{Manipulation of spin}
In the preceding analysis, the MFG sensitivity was described by manipulating the mechanical degrees of freedom and tracing over the spin degrees of freedom. In this section, we show that manipulation of spin degree of freedom can also provide the MFG sensitivity. The essential idea is to prepare the spin-mechanical system in a separable state and allow it to undergo free evolution in the presence of spin-mechanical coupling. The result of this is that mechanical coherent states acquire different phase depending on the spin state \cite{scala2013,wan2016}. For a particular evolution of time, the acquired phase can be completely written on the spin states, which can be read by using Ramsey interferometry \cite{rondin2014}.

\begin{figure}[!ht]
\begin{center}
\includegraphics[scale=0.3]{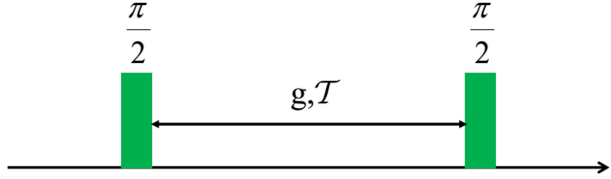}
\caption{Schematic of the Ramsey microwave pulse sequence.}
\label{fig6}
\end{center}
\end{figure} 

Here, we consider a single spin NV center, as shown in Fig. \ref{fig1}, driven by microwave fields and coupled to a mechanical degree of freedom by means of MFG. In diamond substrates, $m_{s}=\pm 1$ levels are typically degenerate. However, this degeneracy can be broken either by a strain field or by applying a static magnetic field. Thus, it is sufficient to use two-levels for the spin manipulation. From now onward, we consider only the $|0\rangle\rightarrow|+1\rangle$ transition of the spin. The Ramsey setup used to manipulate the spin of NV center, is shown schematically in Fig. \ref{fig6}. In the first step, the hybrid system is prepared in a separable state $|\psi\left(0\right)\rangle=|0\rangle|\alpha\rangle$. Here, the mechanical system is assumed to be in coherent state $|\alpha\rangle$. Now, a $\pi/2$ microwave pulse is applied such that spin state is rotated into the superposition state
\begin{align}\label{eq27}
|\psi_{1}\rangle=U_{\pi/2}|\psi\left(0\right)\rangle=\left(\frac{|0\rangle-i|+1\rangle}{\sqrt{2}}\right)|\alpha\rangle\;.
\end{align}
Now, the spin-mechanical interaction is introduced by turning on the MFG and letting the system to evolve freely so as to produce the following state
\begin{align}\label{eq28}
|\psi_{2}\left(t\right)\rangle=U_{t}|\psi_{1}\rangle=\left(\frac{|0\rangle|\alpha\left(t,0\right)\rangle-i|+1\rangle|\alpha\left(t,+1\right)\rangle}{\sqrt{2}}\right)\;,
\end{align}
where, $U_{t}=\exp\left(-\frac{iH_{int}t}{\hbar}\right)$. Note that the spin-dependent mechanical coherent states in Eq. (\ref{eq28}) acquire different phases which are transferred to the spin states after an oscillation period $\mathcal{T}=\frac{2\pi}{\omega_{m}}$ and the resultant state can be written as \cite{scala2013}
\begin{align}\label{eq29}
|\psi_{2}\left(\mathcal{T}\right)\rangle=\left(\frac{|0\rangle-i\exp{\left(\frac{4ig^{2}\mathcal{T}}{\omega_{m}}\right)}|+1\rangle}{\sqrt{2}}\right)|\alpha\rangle\;.
\end{align}
The phase difference written on the spin states can be revealed by applying second $\pi/2$ microwave pulse such that the final state becomes
\begin{align}\label{eq30}
|\psi_{3}\rangle=\frac{1}{2}\left[\left(1-\exp\left(\frac{4ig^{2}\mathcal{T}}{\omega_{m}}\right)\right)|0\rangle-i\left(1+\exp\left(\frac{4ig^{2}\mathcal{T}}{\omega_{m}}\right)\right)|1\rangle\right]|\alpha\rangle\;.
\end{align} 
Thus, the population of ground state $|0\rangle$ can be written as
\begin{align}\label{eq31}
P_{0}=\frac{1}{2}\left(1-\cos\left(\frac{4g^{2}\mathcal{T}}{\omega_{m}}\right)\right)\;.
\end{align}
The population in Eq. (\ref{eq31}) quantifies the following fluorescence signal \cite{pham2013,taylor2008} which can be monitored to yield the MFG sensitivity
\begin{align}\label{eq32}
\mathcal{S}=\frac{\beta_{0}+\beta_{1}}{2}-\frac{\beta_{0}-\beta_{1}}{2}\cos\left(\frac{4g^{2}\mathcal{T}}{\omega_{m}}\right)\;,
\end{align}
where, $\beta_{0}$ is the number of photons collected in one measurement in the absence of spin-mechanical interaction i.e.  the number of photons detected from each spin in $|0\rangle$ state and $\beta_{1}$ is the number of photons collected in one measurement when there is a phase accumulation of $\pi$ during free precision i.e. number of photons detected from each spin in $|+1\rangle$ state. Notably, the performance of the magnetometer is ultimately limited by quantum fluctuations associated with photon shot noise and spin projection noise. Thus, the total noise in the MFG is given by the quadrature sum of these sources of noise \cite{savukov2005}
\begin{align}\label{eq33}
\delta B_{0}=\sqrt{\delta B_{0_{psn}}^{2}+\delta B_{0_{spn}}^{2}}\;,
\end{align}
where $\delta B_{0_{psn}}$ and $\delta B_{0_{spn}}$ represent the contributions from photon-shot noise and spin-projection noise, respectively \cite{itano1993}. In NV-magnetometers, a very small number of photons are collected from each NV spin, which give rise to photon-shot noise. Thus, the photon-shot noise limited uncertainty can be approximated as $\delta B_{0_{psn}}\approx\sqrt{\beta}$, where $\beta=\frac{\beta_{0}+\beta_{1}}{2}$ is the average number of photons collected per measurement. On the other hand, the spin-projection noise arises due to random projection of spin states into one of the states compatible with the measurement process \cite{itano1993} and is given by \cite{fang2013}, $\delta B_{0_{spn}}=\langle\sigma_{z}\rangle=\sqrt{\langle\psi_{3}|\sigma_{z}^{2}|\psi_{3}\rangle-\langle\psi_{3}|\sigma_{z}|\psi_{3}\rangle^{2}}$. Using, Eqs. (\ref{eq30},\ref{eq33}), the minimum uncertainty in the MFG measurement due to the combination of photon-shot noise and spin-projection noise can be written as 
\begin{align}\label{eq34}
\left(\delta B_{0}\right)_{\mbox{min}}=\frac{\sqrt{\beta+1-\cos^{4}\left(\frac{4g^{2}\mathcal{T}}{\omega_{m}}\right)}}{\mbox{max}\bigl|\frac{\partial\mathcal{S}}{\partial B_{0}}\bigr|}\;.
\end{align}
Further, using Eq. (\ref{eq32}), MFG noise becomes
\begin{align}\label{eq35}
\left(\delta B_{0}\right)_{\mbox{min}}=\frac{\hbar\omega_{m}\sqrt{\beta+1-\cos^{4}\left(\frac{4g^{2}\mathcal{T}}{\omega_{m}}\right)}}{8g\mu_{B}g_{l}z_{0}\mathcal{T}\mathcal{C}\beta}\;,
\end{align} 
where, $\mathcal{C}$ is the measurement contrast. Thus, the MFG sensitivity limited by photon-shot noise and spin-projection noise is given by
\begin{align}\label{eq36}
\eta_{B}=\left(\delta B_{0}\right)_{\mbox{min}}\sqrt{t_{m}}\approx\frac{\hbar\omega_{m}\sqrt{\beta+1-\cos^{4}\left(\frac{4g^{2}\mathcal{T}}{\omega_{m}}\right)}}{8g\mu_{B}g_{l}z_{0}\sqrt{\mathcal{T}}\mathcal{C}\beta}\;,
\end{align}
where, the measurement time $t_{m}$ is approximated by $\mathcal{T}$. Note that the MFG sensitivity in Eq. (\ref{eq36}) depends on the number of photons collected per measurement $\beta$, resonance contrast $\mathcal{C}$, measurement time, and the spin-mechanical coupling.
\begin{figure}[!ht]
\begin{center}
\includegraphics[scale=0.6]{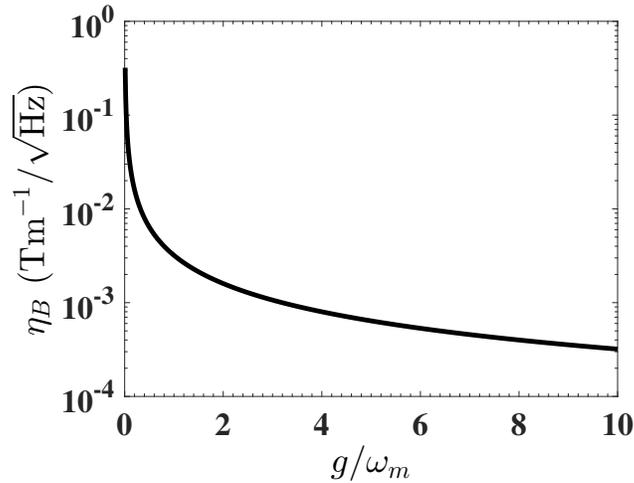}
\caption{MFG sensitivity versus scaled spin-mechanical coupling. Parameters chosen are $\mathcal{C}$=5$\%$, $\beta$=10 kcount s$^{-1}$, and $\omega_{m}=$38 kHz. }
\label{fig7}
\end{center}
\end{figure}
For an observed contrast of 5$\%$ and count rate of 10 kcount s$^{-1}$ \cite{levi2015}, the MFG sensitivity is shown in Fig. \ref{fig7}. It can be inferred from Fig. \ref{fig7} that increase in spin-mechanical coupling results in decrease in MFG sensitivity with a value reaching upto 100 $\mu$Tm$^{-1}$/$\sqrt{\mbox{Hz}}$ at $g=10\omega_{m}$. Further, we find that the above analysis remains unaffected even if the initial mechanical coherent state is replaced by a thermal state \cite{scala2013,wan2016}.

\subsection{Experimental accessibility} 

In the above discussion, we did not include the decoherence from the mechanical oscillator as the decoherence times are substantially longer than the free evolution time $\mathcal{T}$. Thus, we can neglect the mechanical decoherence. In our analysis above, we considered a nanodiamond of radius 50 nm optically levitated in a dipole trap in ultrahigh vacuum along with feedback cooling \cite{levi2015} and oscillating at a mechanical frequency $\omega_{m}=38$ kHz. For such a configuration, the decoherence time corresponding to gas damping is of the order of $T_{g}\sim$10$^{6}$ s which is quite large as compare to $\mathcal{T}$ ($\sim$200 $\mu$s). So, the decoherence due to gas damping can be neglected. Further, the 
feedback induced decoherence  \cite{levi2015} operates over a longer timescale ($T_{fb}\sim$ ms) than the free evolution time (200 $\mu$s) and its effect can also be neglected. Furthermore, in our model, the decoherence due to optical scattering ($T_{opt}=\gamma_{opt}^{-1}\sim$ ms) can also be ignored for the duration $\mathcal{T}\sim$200 $\mu$s. Other sources of decoherence include the dephasing of the NV center due to interaction with other (non-NV) spins in the diamond lattice. However, the associated dephasing time ($\sim$ 400 ms \cite{naydenov2011}) is still larger so as not to affect the MFG sensitivity.

\section{Conclusion}
In conclusion, an NV-center based hybrid nanomechanical magnetometer is proposed where a spin degree of freedom is coupled to a mechanical degree of freedom by an MFG. We have described two schemes for MFG sensing: first by analyzing the position spectrum of the mechanical oscillator, and second by manipulating the spin degrees of freedom. The first procedure gives a MFG sensitivity of 1 $\mu$Tm$^{-1}$/$\sqrt{\mbox{Hz}}$ whereas the latter provides 100 $\mu$Tm$^{-1}$/$\sqrt{\mbox{Hz}}$. We have explained in detail 
that such an experimentally accessible nanomagnetometer functions well under both cooling and ambient conditions and thereby provides a platform for sensitive magnetometry applications. Our scheme can be extended to performing vector magnetic gradiometry by coupling the three independent directions of nanodiamond oscillations to the three cartesian components of MFG.

\section*{Funding}
This research was funded by the Office of Naval Research (ONR) under Award No. N00014-14-1-0803.

\end{document}